\documentclass[12pt]{article}
\usepackage{graphicx}
\def\hybrid{\topmargin 0pt      \oddsidemargin 0pt
        \headheight 0pt \headsep 0pt
        \voffset=-0.5cm
        \textwidth 6.25in       
        \textheight 9.5in       
        \marginparwidth 0.0in
        \parskip 5pt plus 1pt   \jot = 1.5ex}
\catcode`\@=11
\def\marginnote#1{}

\newcount\hour
\newcount\minute
\newtoks\amorpm
\hour=\time\divide\hour by60
\minute=\time{\multiply\hour by60 \global\advance\minute by-\hour}
\edef\standardtime{{\ifnum\hour<12 \global\amorpm={am}%
        \else\global\amorpm={pm}\advance\hour by-12 \fi
        \ifnum\hour=0 \hour=12 \fi
        \number\hour:\ifnum\minute<10 0\fi\number\minute\the\amorpm}}
\edef\militarytime{\number\hour:\ifnum\minute<10 0\fi\number\minute}

\def\draftlabel#1{{\@bsphack\if@filesw {\let\thepage\relax
   \xdef\@gtempa{\write\@auxout{\string
      \newlabel{#1}{{\@currentlabel}{\thepage}}}}}\@gtempa
   \if@nobreak \ifvmode\nobreak\fi\fi\fi\@esphack}
        \gdef\@eqnlabel{#1}}
\def\@eqnlabel{}
\def\@vacuum{}
\def\draftmarginnote#1{\marginpar{\raggedright\scriptsize\tt#1}}
\def\draftlabel#1{{\@bsphack\if@filesw {\let\thepage\relax
   \xdef\@gtempa{\write\@auxout{\string
      \newlabel{#1}{{\@currentlabel}{\thepage}}}}}\@gtempa
   \if@nobreak \ifvmode\nobreak\fi\fi\fi\@esphack}
        \gdef\@eqnlabel{#1}}
\def\@eqnlabel{}
\def\@vacuum{}
\def\draftmarginnote#1{\marginpar{\raggedright\scriptsize\tt#1}}

\def\draft{\oddsidemargin -.5truein
        \def\@oddfoot{\sl preliminary draft \hfil
        \rm\thepage\hfil\sl\today\quad\militarytime}
        \let\@evenfoot\@oddfoot \overfullrule 3pt
        \let\label=\draftlabel
        \let\marginnote=\draftmarginnote
   \def\@eqnnum{(\theequation)\rlap{\kern\marginparsep\tt\@eqnlabel}%
\global\let\@eqnlabel\@vacuum}  }

\def\numberbysection{\@addtoreset{equation}{section}
        \def\theequation{\thesection.\arabic{equation}}}

\def\underline#1{\relax\ifmmode\@@underline#1\else
        $\@@underline{\hbox{#1}}$\relax\fi}

\def\titlepage{\@restonecolfalse\if@twocolumn\@restonecoltrue\onecolumn
     \else \newpage \fi \thispagestyle{empty}\c@page\z@
        \def\thefootnote{\fnsymbol{footnote}} }

\def\endtitlepage{\if@restonecol\twocolumn \else  \fi
        \def\thefootnote{\arabic{footnote}}
        \setcounter{footnote}{0}}  
\relax

\hybrid

\newfont{\Bbb}{msbm10 scaled 1\@ptsize00}

\newfont{\Bbbb}{msbm7 scaled 1\@ptsize00}
\newcommand{\cc}{\raise-1pt\hbox{$\mbox{\Bbbb C}$}}
\newcommand{\zz}{\raise-1pt\hbox{$\mbox{\Bbbb Z}$}}

\def\beq{\begin{equation}}
\def\eeq{\end{equation}}
\def\p{\partial}

\def\DD{{\sf D}}

\def\lbracket{\Bigl <}
\def\rbracket{\Bigr >}

\begin{document}

\begin{titlepage}

\title{Dyson diffusion on a curved contour}

\author{
A.~Zabrodin\thanks{
Skolkovo Institute of Science and Technology, 143026, Moscow, Russia and
National Research University Higher School of Economics,
20 Myasnitskaya Ulitsa,
Moscow 101000, Russia and
NRC KI, KCTEP, Moscow, Russia;
e-mail: zabrodin@itep.ru}}

\date{November 2022}
\maketitle

\vspace{-7cm} \centerline{ \hfill ITEP-TH-25/22}\vspace{7cm}

\begin{abstract}

We define the Dyson diffusion process on a curved smooth closed contour
in the plane and derive the Fokker-Planck equation for probability density.
Its stationary solution is shown to be
the Boltzmann weight for the logarithmic gas confined on the contour.

\end{abstract}


\end{titlepage}

\section{Introduction}

In the pioneering paper \cite{Dyson1} Dyson defined a diffusion 
process for eigenvalues of a unitary matrix confined to the unit circle. 
In the present paper we introduce a more general diffusion process for $N$ 
interacting particles in an external potential
confined to a smooth closed contour of arbitrary shape in the plane. 
This is done with the help of $N$ copies 
of the standard Brownian motion $B_i(t)$. 

The time evolution of the probability density 
is described by the Fokker-Planck 
equation. We derive this equation and find its stationary solution
which is the Boltzmann weight for the logarithmic gas in the external
potential. 
The logarithmic gases were introduced by Dyson in papers 
\cite{Dyson}, where it was shown that eigenvalues of random matrices can be represented
as a statistical ensemble of charged particles with 2D Coulomb (logarithmic) interaction. A comprehensive presentation of the theory of logarithmic gases
can be found in \cite{Forrester}. In our case we obtain the Boltzmann 
weight for the logarithmic gas on the curved contour at inverse 
temperature $\beta$ studied in \cite{WZ22}. In section \ref{section:free}
we present the results for the large $N$ expansion of the free energy
of this gas.

\section{The diffusion process on a curved contour}

Let $\Gamma$ be a smooth non-self-intersecting closed 
contour in the plane. We consider a system of $N$ interacting particles 
with complex coordinates $z_i \in \Gamma$ moving along the 
contour and subject to Gaussian random forces. 

To describe the diffusion process, we introduce $N$ copies 
of the standard Brownian motion $B_i(t)$ such that
$B_i(0)=0$ satisfying the conditions
\beq\label{d1}
\lbracket B_i(t)\rbracket =0, \qquad
\lbracket (B_i(t)-B_i(t'))(B_j(t)-B_j(t'))\rbracket =\delta_{ij}
|t-t'|,
\eeq
or
$$
\lbracket \dot B_i(t)\dot B_j(t')\rbracket =\delta_{ij}\delta (t-t').
$$
For $dB_i(t)=B_i(t+dt) -B_i(t)$ we have
\beq\label{d2}
\lbracket dB_i(t)\rbracket =0, \qquad 
\lbracket dB_i(t)dB_j(t)\rbracket =0 \quad \mbox{for $i\neq j$}, \qquad
\lbracket (dB_i(t))^2\rbracket =dt,
\eeq
so we should consider $(dB_i)^2$ as a first order quantity in $dt$. 

Below we use the standard notions and facts from stochastic calculus
and the theory of stochastic differential equations \cite{book1,book2}.
We can write
$$
dz_i(t)=z_i(t+dt)-z_i(t), 
$$
where both points $z_i(t)$ and $z_i(t+dt)$ belong to $\Gamma$. 
For $z\in \Gamma$, let $\tau (z)$ and $\nu (z)$ be respectively 
the unit tangential and normal vectors to $\Gamma$ at the point $z$ 
represented as complex numbers. We assume that the tangential vector 
looks in the counterclockwise direction and the normal
vector is directed to the exterior of the contour, so we have $\tau =i\nu$.
For $z=z_i$ we will write
$\tau (z_i)=\tau_i$, $\nu (z_i)=\nu_i$ for brevity.
Then $\tau_i |dz_i|$ is the tangent vector at $z_i$ with length $|dz_i|$.
However, for the needs of stochastic calculus we can not
write simply $dz_i=\tau_i |dz_i|$ because it is necessary 
to expand up to the
second order. It is easy to see that the second order correction is 
$-\nu_i |dz_i|\frac {d\vartheta_i}{2}$, where $\vartheta (z)=
\mbox{arg}\, \tau (z)$. By definition, $k(z)=d\vartheta (z)/|dz|$
is the curvature of the curve $\Gamma$ at the point $z$, so we write
\beq\label{d3}
dz_i=\tau_i |dz_i| -\frac{1}{2}\, \nu_i k_i |dz_i|^2.
\eeq

We define the diffusion process by setting
\beq\label{d4}
|dz_i|=\p_{s_i}E dt +\sqrt{\kappa}dB_i.
\eeq
Here
\beq\label{d5}
E=2\sum_{i<j}\log |z_i-z_j| + \sum_i W(z_i),
\eeq
$W(z)$ is an external potential, $\p_s =
\tau \p_z +\bar \tau \p_{\bar z}$ means the tangential
derivative and $\kappa$ is a parameter characterizing strength 
of the random force. The real-valued function $W$ is defined
in a strip-like neighborhood of the contour and depends on both 
$z$ and $\bar z$
(we write $W=W(z)$ just for simplicity of the notation).
Explicitly, we have:
\beq\label{d6}
\p_{s_i}E =\tau_i \sum_{j\neq i}\frac{1}{z_i-z_j}+\tau_i \p_{z_i}W(z_i) 
+\mbox{c.c.}
\eeq
Substituting (\ref{d4}) into equation (\ref{d3}), we define the 
diffusion process on $\Gamma$ as the following stochastic differential
equation:
\beq\label{d7}
dz_i=\tau_i \p_{s_i}E dt -\frac{\kappa}{2}\, \nu_i k_i (dB_i)^2 +
\sqrt{\kappa} \tau_i dB_i.
\eeq
In the right hand side, we keep only terms which are of the first
order in $dt$. 

Let us note that equation (\ref{d7}) agrees with the familiar diffusion 
process on the unit circle, where $z_j =e^{i\theta_j}$
(see \cite{Dyson1}). In terms of $\theta_j$'s, the stochastic 
differential equation (the Langevin equation) is of the standard form
\beq\label{d8}
d\theta_j =\p_{\theta_j}E dt +\sqrt{\kappa}dB_j.
\eeq
Applying the It\^o formula 
$$
df(\theta_j )=\frac{\p f}{\p \theta_j} \, d\theta_j +
\frac{\kappa}{2}\, \frac{\p^2 f}{\p \theta_j^2}\, (dB_j)^2
$$
to the function $f(\theta )=e^{i\theta}$, we can rewrite equation 
(\ref{d8}) in the form
\beq\label{d9}
dz_j = iz_j \p_{\theta_j}E dt-\frac{\kappa}{2} z_j (dB_j)^2 +i
z_j \sqrt{\kappa} dB_j
\eeq
which agrees with (\ref{d7}) since for the unit circle $\tau_j =iz_j$ and
$k_j=1$. Equation (\ref{d7}) is a generalization of (\ref{d9}) for 
curved contours of arbitrary shape.

Let us now derive a stochastic differential equation for an arbitrary
(smooth enough) function $f=f(z_1, \ldots , z_N)$. As is customary in
the It\^o calculus, the differential $df$ 
should contain not only first order terms in $dz_i$ but also 
the second order terms:
\beq\label{d10}
df = \sum_i \Bigl (\frac{\p f}{\p z_i}\, dz_i +
\frac{\p f}{\p \bar z_i}\, d\bar z_i\Bigr )+\frac{1}{2} \sum_i
\Bigl (\frac{\p^2 f}{\p z_i^2}\, (dz_i)^2 + 
\frac{\p^2 f}{\p \bar z_i^2}\, (d\bar z_i)^2 +2
\frac{\p^2 f}{\p z_i \p \bar z_i}\, |dz_i|^2 \Bigr ).
\eeq
Using (\ref{d7}) and taking into account that $\nu_i \p_{z_i}f +
\bar \nu_i \p_{\bar z_i}f=\p_{n_i}f$ 
is the normal derivative of the function $f$ at the point $z_i \in \Gamma$
(the normal vector looks to the exterior of $\Gamma$), we have from
(\ref{d10}):
\beq\label{d11}
\begin{array}{l}
\displaystyle{
df=\sum_i \p_{s_i}E \p_{s_i}f dt +\frac{\kappa}{2}
\sum_i \Bigl (-k_i \p_{n_i}f + \tau_i^2 \p^2_{z_i}f +
\bar \tau_i^2 \p^2_{\bar z_i}f +2\p_{z_i}\p_{\bar z_i}f \Bigr )(dB_i)^2}
\\ \\
\displaystyle{
\phantom{aaaaaaaaaaaaaaaaaaaaaaaaaaaaaaaaaaaa}
+\sqrt{\kappa} \sum_i \p_{s_i}f \, dB_i}.
\end{array}
\eeq
According to the rules of It\^o calculus, one should put
$(dB_i)^2 =dt$. In this way, we obtain the stochastic differential
equation
\beq\label{d12}
df=\sum_i \left (\vphantom{\frac{A}{B}}\p_{s_i}E \p_{s_i}f
+\frac{\kappa}{2} \Bigl (-k_i \p_{n_i}f + \tau_i^2 \p^2_{z_i}f +
\bar \tau_i^2 \p^2_{\bar z_i}f +2\p_{z_i}\p_{\bar z_i}f \Bigr )\right )dt
+\sqrt{\kappa} \sum_i \p_{s_i}f \, dB_i.
\eeq
In order to simplify the right hand side, 
we write for a function $f=f(z)$:
$$
\p_s^2f=\p_s(\tau \p_z f)+\p_s (\bar \tau \p_{\bar z}f)=
\p_s \tau \cdot \p_zf +\tau \p_s (\p_z f)\, +\, \mbox{c. c.}
$$
Taking into account that $\p_s \tau = i\tau k$, we have:
$$
\p_s^2f=-k (\nu \p_z f +\bar \nu \p_{\bar z}f)+\tau (
\tau \p^2_z f +\bar \tau \p_z \p_{\bar z}f) + \bar \tau (
\tau \p_z \p_{\bar z}f +\bar \tau \p^2_{\bar z}f)
$$
or, finally,
\beq\label{d13}
\p_s^2f=-k\p_n f +\tau^2 \p^2_z f +\bar \tau^2 \p^2_{\bar z}f +
2 \p_z\p_{\bar z}f.
\eeq
Substituting this into the right hand side of (\ref{d12}), we obtain the 
stochastic differential equation in the simple form
\beq\label{d14}
df=\sum_i \left (\frac{\kappa}{2}\, \p_{s_i}^2 f +\p_{s_i}E \p_{s_i}f\right )
dt + \sqrt{\kappa} \sum_i \p_{s_i}f \, dB_i,
\eeq
with the derivatives in the tangential direction only. 

\section{The Fokker-Planck equation and its stationary solution}

Let $\hat A$ be the differential operator
\beq\label{f1}
\hat A=\sum_i \left (\frac{\kappa}{2}\, \p_{s_i}^2  +
\p_{s_i}E \p_{s_i}\right ),
\eeq
then equation (\ref{d14}) can be written as
$$
df=\hat A f +\sqrt{\kappa} \sum_i \p_{s_i}f \, dB_i.
$$
As is known, the Fokker-Planck  
equation for the time 
evolution of the probability density
$P=P(z_1, \ldots , z_N)$ is 
$$
\p_t P =\hat A^{*}P,
$$
where $\hat A^{*}$ is the operator adjoint to $\hat A$. 
Explicitly, in our case the Fokker-Planck equation is 
\beq\label{f2}
\p_t P=\sum_i \left (\frac{\kappa}{2}\, \p^2_{s_i}P-\p_{s_i}E \p_{s_i}P
-\p^2_{s_i}E P\right ).
\eeq

The similarity transformation of the operator 
$\hat A^{*}$ with the function $e^{E/\kappa}$ 
eliminates the first order derivatives. With the help of this procedure,
we obtain
the ``Hamiltonian'' $\hat H$:
\beq\label{H1}
\hat H = e^{-E/\kappa} \, \hat A^{*} \, e^{E/\kappa} =
\frac{1}{2}\sum_i \left ( \kappa \p_{s_i}^2 -\kappa^{-1} (\p_{s_i}E)^2
-\p_{s_i}^2 E\right ).
\eeq
Note that if the contour $\Gamma$ is the unit circle and $W=0$, 
then $\hat H$ is the Hamiltonian 
of the Calogero-Sutherland model:
\beq\label{cs}
\hat H=\frac{\kappa}{2}\left (\sum_i \p^2_{\theta_i} -
\frac{2}{\kappa}\Bigl (\frac{2}{\kappa}-1\Bigr ) \sum_{i\neq j}
\frac{1}{4\sin^2((\theta_i -\theta_j)/2)}\right ) +\mbox{const}.
\eeq

The stationary solution of (\ref{f2}), $P_0$, 
such that $\p_t P_0=0$ is the zero
mode of the operator $\hat A^{*}$, $\hat A^{*}P_0=0$. 
It is easy to check that it has the form
\beq\label{f3}
P_0=\exp \Bigl (\frac{2}{\kappa}\, E\Bigr ).
\eeq
Denoting
\beq\label{f4}
\beta =\frac{2}{\kappa}
\eeq
and substituting equation (\ref{d5}) for $E$, we have:
\beq\label{f5}
P_0(z_1, \ldots , z_N)=\exp \left (2\beta \sum_{i<j}\log |z_i-z_j|
+\beta \sum_i W(z_i)\right )=
\prod_{i<j}|z_i-z_j|^{2\beta}\prod_{k}e^{\beta W(z_k)}.
\eeq
This is the Boltzmann weight for the logarithmic gas in the external
potential $W$ at the inverse temperature $\beta$. 
The Boltzmann weight (\ref{f5}) defines the $\beta$-ensemble
of $N$ particles on a simple closed contour
$\Gamma$ in the plane \cite{WZ22}. 
An early reference on this problem is \cite{Jancovici}
(section 7). A similar problem having support in the complex plane, 
the 2D Dyson gas, had been considered in \cite{WZ03,WZ03a,WZ06,Z04}.

\section{Free energy of the logarithmic gas: large $N$
asymptotics}
\label{section:free}

For completeness, we present here the results for the large $N$ 
asymptotics of the free energy of the logarithmic gas obtained in \cite{WZ22}
(see also \cite{Johansson21} for the rigorous proof for $\beta =1$).
We also give a new interpretation of the $O(1)$-contribution. 

The partition function is the $N$-fold integral
\beq\label{a1}
Z_N= \oint_{\Gamma}\ldots \oint_{\Gamma}P_0(z_1, \ldots , z_N)
ds_1 \ldots ds_N,
\eeq
with $P_0$ given by (\ref{f5}),
where $ds_i=|dz_i|$ is the line element along the contour. 
As $N\to \infty$, the partition function is known to behave as
$$
Z_N =N! N^{(\beta -1)N}e^{F^{(N)}},
$$
where the free energy $F_N$ has the expansion in integer powers of $N$
of the form
\beq\label{a2}
F^{(N)}=N^2 F_0 + N F_1 +F_2 + O(1/N).
\eeq
The non-vanishing contributions as $N\to \infty$, $F_0, F_1$ and $F_2$,
are of primary interest. 

The curve $\Gamma$ divides the plane into the interior domain, 
$\DD _{\rm int}$, and the exterior domain, $\DD _{\rm ext}$,
containing $\infty$. 
Without loss of generality, we assume that $0\in \DD _{\rm int}$.
The results for $F_0, F_1, F_2$ are expressed in terms of conformal
maps $w_{\rm int}(z)$,
$w_{\rm ext}(z)$ 
of the domains $\DD_{\rm int}$, $\DD_{\rm ext}$ to the interior
and exterior of the unit circle respectively. We normalize the 
conformal maps as follows:
\beq\label{a3}
\begin{array}{l}
w_{\rm int}(0)=0, \qquad w_{\rm int}'(0)>0,
\\ \\
w_{\rm ext}(\infty )=\infty , \qquad w_{\rm ext}'(\infty )=1/r >0.
\end{array}
\eeq
The quantity $r$ is called the conformal radius of the 
domain $\DD _{\rm ext}$.
It is also convenient to use the notation
\beq\label{a3a}
\psi_{\rm int}(z)=\log |w_{\rm int}'(z)|, \qquad
\psi_{\rm ext}(z)=\log |w_{\rm ext}'(z)|.
\eeq
Note that $\psi_{\rm int}$, $\psi_{\rm ext}$ are harmonic functions 
in the domains $\DD_{\rm int}$, $\DD_{\rm ext}$ respectively.

In \cite{WZ22} it was shown that
the leading and sub-leading contributions to the free energy
are given by
\beq\label{a4}
\begin{array}{l}
F_0=\beta \log r,
\\ \\
\displaystyle{
F_1=\frac{\beta}{2\pi}\oint_{\Gamma}
|w_{\rm ext}'|Wds -(\beta -1) \log \frac{er}{2\pi \beta} +
\log \frac{2\pi}{\Gamma (\beta )}}.
\end{array}
\eeq
The result for $F_2$ is most interesting and non-trivial. 
The contribution
$F_2$ can be naturally 
divided into two parts, ``classical'' and ``quantum'': 
$F_2= F_2^{(cl)}+F_2^{(q)}$. The ``classical'' part has the electrostatic 
nature while the ``quantum'' part is due to fluctuations of the particles.
The result for the ``classical'' part is expressed through the 
Neumann jump operator $\hat {\cal N}$ associated with the contour.
This operator takes a function $f$ on $\Gamma$ to the jump of normal
derivative of its harmonic continuations $f_H$ to $\DD_{\rm int}$ and 
$f^H$ to $\DD_{\rm ext}$:
$$
\hat {\cal N}f = \p_nf_H -\p_n f^H,
$$
where the normal vector looks to the exterior of $\Gamma$ in both cases.
The result for $F_2^{(cl)}$ is as follows:
\beq\label{a5}
F_2^{(cl)}=\frac{\beta}{8\pi}\oint_{\Gamma}\Bigl ( (1-\beta^{-1})
\psi_{\rm ext} +W\Bigr ) \hat {\cal N}\Bigl ((1-\beta^{-1})
\psi_{\rm ext} +W\Bigr )ds.
\eeq
The result for $F_2^{(q)}$ is
\beq\label{a6}
F_2^{(q)}=\frac{1}{24\pi}\oint_{\Gamma} \Bigl (
\psi_{\rm int}\p_n \psi_{\rm int} -\psi_{\rm ext}\p_n \psi_{\rm ext}
\Bigr )ds +\frac{1}{6}\Bigl (\psi_{\rm ext}(\infty )-
\psi_{\rm int}(0)\Bigr )+\log \sqrt{\beta}.
\eeq
Remarkably, the quantity in the right hand side 
of (\ref{a6}) is $\frac{1}{24}$ 
times the
Loewner energy $I^L(\Gamma )$ of the curve $\Gamma$:
\beq\label{a7}
F_2^{(q)}=\frac{1}{24}\, I^L(\Gamma )+\log \sqrt{\beta},
\eeq
where
\beq\label{a8}
I^L (\Gamma )=\frac{1}{\pi}\int_{\DD _{\rm int}} |\nabla \psi_{\rm int}|^2
d^2 z + \frac{1}{\pi}\int_{\DD _{\rm ext}} |\nabla \psi_{\rm ext}|^2
d^2 z + 4\Bigl (\psi_{\rm ext}(\infty )-\psi_{\rm int}(0)\Bigr )
\eeq
(see \cite{Wang}). The notion of Loewner energy was recently actively
discussed in the literature in the context of the Schramm-Loewner evolution
(SLE), see, e.g., the survey \cite{Wang1} and references therein. 
The quantity (\ref{a8}) is also known as the universal
Liouville action \cite{TT}.

\section{Concluding remarks}

In this paper we have defined the diffusion process for $N$ interacting
particles on a smooth closed contour $\Gamma$ in the plane. Each particle
is subject to action of a random force represented by the 
Brownian process $B(t)$ such that
$\lbracket (dB(t))^2\rbracket =\kappa dt$, where the coefficient
$\kappa$ characterizes strength of the random force. We have derived the 
Fokker-Planck equation for the corresponding 
probability density. The stationary solution
of the Fokker-Planck equation was shown to be given by the Boltzmann 
weight for the $N$ particles confined on the contour 
interacting via the 2D Coulomb (logarithmic)
potential. The temperature of this logarithmic gas is $\kappa /2$.
This model was studied in the paper \cite{WZ22}, where the 
non-vanishing contributions to the free energy as $N\to \infty$
were found. The leading term is of order
$O(N^2)$ and it has the electrostatic nature.
The most interesting quantity is the ``quantum'' part
(i.e., the part which is entirely due to fluctuations) of the 
$O(1)$-contribution to the free energy. Although the result 
(\ref{a6}) for it was already 
obtained in \cite{WZ22}, in that paper it was not
noticed that it is actually $\frac{1}{24}$ of the Loewner energy 
$I^L(\Gamma )$ of the contour $\Gamma$. We make this explicit 
in the present paper. 

\section*{Acknowledgments}


The author thanks P. Wiegmann for useful discussions and bringing
the paper \cite{Wang1} to his attention.

\end{document}